\documentclass[%
 preprint, 3p, twocolumn,
]{elsarticle}

\usepackage{graphicx}
\usepackage{dcolumn}
\usepackage{bm}
\usepackage{float} 
\usepackage{ulem} 
\usepackage{xcolor}
\usepackage{amsmath}
\usepackage{hyperref}
\usepackage[mathlines]{lineno}

\begin{document}

\title{New Constraints on Supersymmetry Using Neutrino Telescopes}
\author[1]{Meighen-Berger S.\corref{cor1}\fnref{fn1}}
\author[2,1]{Agostini M.}
\author[1]{Ibarra A.}%
\author[1]{Krings K.}%
\author[1]{Niederhausen H.}
\author[1]{Rappelt A.}%
\author[1]{Resconi E.}%
\author[1]{Turcati A.}

\cortext[cor1]{Corresponding author}
\fntext[fn1]{stephan.meighen-berger@tum.de}

\address[1]{Technische Universit\"at M\"unchen, James-Franck-Stra{\ss}e, 85748, Garching, Germany}
\address[2]{Department of Physics and Astronomy, University College London, Gower Street, London WC1E 6BT, UK}

\date{\today}

\begin{abstract}
We demonstrate that megaton-mass neutrino telescopes are able to observe the signal from long-lived particles beyond the Standard Model, in particular the stau, the supersymmetric partner
of the tau lepton. Its signature is an excess of charged particle tracks with horizontal arrival directions and energy deposits between 0.1 and 1\,TeV inside the detector. We exploit this previously-overlooked signature to search for stau particles in the publicly available IceCube data. The data shows no evidence of physics beyond the Standard Model. We derive a new lower limit on the stau mass of $320$\,GeV (95\% C.L.) and estimate that this new approach, when applied to the full data set available to the IceCube collaboration, will reach word-leading sensitivity to the stau mass ($m_{\tilde{\tau}}=450\,\mathrm{GeV}$).
\end{abstract}

\maketitle


New long-lived particles are an integral part of many theories beyond the Standard Model (SM). Supersymmetry, for example, predicts the existence of the stau, the supersymmetric partner of the tau lepton. The stau is long-lived in scenarios in which the gravitino is the lightest among all supersymmetric partners, and the stau is the next-to-lightest. In this case, and provided R-parity is conserved,
the stau can only decay into a gravitino and a tau lepton. The width of this decay is suppressed
by the scale of supersymmetry breaking (for a review, see
\cite{Giudice:1998bp}). As a result, the stau lifetime can be as long as several
seconds, minutes or even years, depending on the model parameters.
The most sensitivity searches for the stau have been performed at the Large Hadron Collider by the ATLAS and CMS collaborations. Using the mass of the stau ($m_{\widetilde\tau}$) as a free parameter, they reported $m_{\widetilde\tau}\geq 430$ GeV
and $m_{\widetilde\tau}\geq 240$ GeV at $95\%\,\mathrm{C.L.}$ respectively \cite{Aaboud:2019trc}. For these limits in particular, the stau's mass is the only parameter of interest, due to the assumed Drell-Yan production.

\begin{figure}[t]
   \centering
   \includegraphics[scale=1]{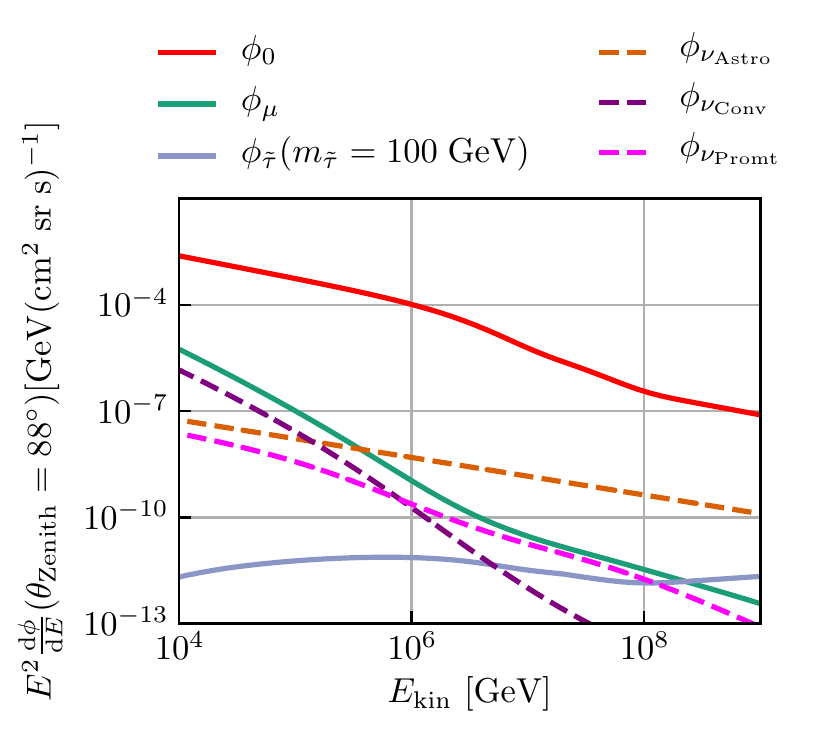}
   \caption{Predicted cosmic ray flux, $\phi_0$ (red), muon flux $(\phi_\mu$ (green), muon neutrino flux $\phi_\nu$ (dashed) and stau flux $\phi_{\tilde{\tau}}$ (blue) at the surface above the IceCube detector at $\theta_\mathrm{Zenith} = 88^\circ$. The neutrino flux is further split into its components, the astrophysical flux (orange), conventional flux (purple) and prompt flux (magenta). The stau mass was set here to 100 GeV and the interaction model Sibyll 2.3c \cite{Riehn:2017mfm} and primary model and H4a \cite{Gaisser:2011cc} were used. The low stau flux, when compared to muon and neutrino fluxes, makes searches using neutrino telescopes challenging.}
   \label{fig:Flux}
 \end{figure}

Stau searches have also been proposed in the context of megaton-mass neutrino telescopes. Highly energetic cosmic particles (cosmic rays and neutrinos) colliding with nucleons in the Earth's atmosphere are capable of producing staus. These in turn would then appear as charged particles in the detectors. In particular, they would appear as charged particles propagating through the detector, so-called tracks. For such events, other particles producing charged tracks would act as a background, mainly atmospheric muons and muons produced by neutrinos. Figure \ref{fig:Flux} shows the relative fluxes at the surface. The orders of magnitude difference between the stau flux and background makes disentangling them difficult, even for a low stau mass of 100 GeV. The neutrino flux is divided into its primary contributors. These are the astrophysical flux, the conventional flux from the decay of $\pi$ and $K$ mesons and the prompt flux from the decay of heavier mesons.

One proposed strategy to disentangle staus from the background is to search for stau pairs. Due to the highly relativistic boosting, the two staus would move in the same
direction, cross the detector simultaneously, and thus give rise to two parallel tracks --- a smoking-gun signature.  \cite{Albuquerque:2003mi, Huang:2006ie, Albuquerque:2006am,Ahlers:2006pf,Reno:2005si, Albuquerque:2009vk}. However, current telescopes can not distinguish these events from the overwhelming background, single tracks from cosmic ray muons ($hN\rightarrow\mu X$) and/or from charged-current muon neutrino interactions ($\nu N\rightarrow\mu X$), unless the two stau tracks have a large separation (in IceCube by $\sim 135$ meters \cite{Abbasi:2012kza, Soldin:2018vak}). Thus, the majority ($\sim 99.9\,\%$) of the potential stau particles would go undetected.

In this letter we discuss how neutrino telescopes, in particular IceCube, can exploit a different
signature to observe, on a statistical basis, a stau signal. At the energies of interest, staus are expected to be significantly more penetrating than muons of the same energy, because they essentially do not loose energy through stochastic processes \cite{Reno:2005si}. Hence, for nearly horizontal trajectories, tens of kilometers of ice shield IceCube from the vast majority of muons but not from staus. We thus search for an excess of track events over the background expected from muons crossing the detector horizontally at a zenith angles near $\sim\,90^\circ$. In contrast to previous works, this new analysis does not rely on an identifiable double track signature and thus less stringent event selection criteria. We demonstrate the potential strength of the method by analyzing one year of publicly available IceCube data \cite{IceCubeData2018, Aartsen:2016oji}. In order to identify this stau component, we utilize differences in the corresponding energy and angular distributions compared to the contributions from muons.

Considering state-of-the-art cosmic-ray flux and hadronic interaction models, we calculate the stau signal and backgrounds using MCEq~\cite{Fedynitch:2015zma}, a tool that solves the cascade equations that govern the interaction of cosmic rays and propagation of the resulting air shower. We modified MCEq to include the
generation of staus with a probability of~\cite{Ahlers:2007js}
 \begin{equation}
   P_{\widetilde\tau}^\mathrm{h}(E)\approx\frac{A\sigma^\mathrm{h,nucleon}_{\widetilde\tau}}{\sigma^\mathrm{h,air}_\mathrm{total}}.
 \end{equation}
where $\sigma^\mathrm{h,nucleon}_{\widetilde\tau}$ is the total stau production cross-section from the collisions of a hadron $h$ with a nucleon in
the atmosphere, $\sigma^\mathrm{h,air}_\mathrm{total}$ is the total cross-section of $h$ with air, and $A = 14.6$ is the average number of nucleons in a nucleus of air. 
The stau production cross-section has been computed assuming the Drell-Yan process through MadGraph \cite{Alwall:2014hca, Frederix:2018nkq}. More details about these calculations and corresponding assumptions are given in \ref{sec:app:stau_cross}. After production, staus are assumed to propagate straight through matter with
an energy loss per column density, $X$, that can be approximated by:
\begin{equation}\label{eq:energyloss}
  -\frac{\mathrm{d}E}{\mathrm{d}X} = a_{\widetilde\tau}(E) + b_{\widetilde\tau}(E)E\;,
\end{equation}
where $a_{\widetilde\tau}(E)$ accounts for ionization losses and  $b_{\widetilde\tau}(E)$ for energy loss by pair production, hadronic interactions, and bremsstrahlung.
Given that the ionization effects for  stau are expected to be similar to those of a muon we assume $a_{\widetilde\tau}(E)\approx a_\mu(E)$. In contrast,
we expect all other effects to depend on the particle speed and hence to be in good approximation given by $b_{\widetilde\tau}(E)\approx b_\mu(E)
m_\mu/m_{\widetilde\tau}$~\cite{Ahlers:2007js}. We take the parameters related to the muons from Refs. \cite{Tanabashi:2018oca, Groom:2008zz, Groom:2001kq}.

Our backgrounds are muons that produce a detector response indistinguishable from that of the staus. The muon backgrounds can be divided into two components according to whether muons are produced by a hadronic interaction in a cosmic ray airshower or by a neutrino interacting in the Earth. We simulate these contributions similarly to what we do for the staus. In addition to the cosmic-ray flux and composition used before, we now also include the flux of astrophysical neutrinos measured by
IceCube~\cite{Aartsen:2020aqd}:
\begin{equation}
\frac{\mathrm{d}\phi}{\mathrm{d}E}=1.66_{-0.27}^{+0.25}
\left(\frac{E}{100\;\mathrm{TeV}}\right)^{-2.53\pm 0.07}\;.
\end{equation}
To calculate event rates from particle fluxes we rely on publicly available IceCube effective areas, as discussed in \ref{sec:app:effective_areas}.

\begin{figure}[t]
   \centering
   \includegraphics[scale=1.]{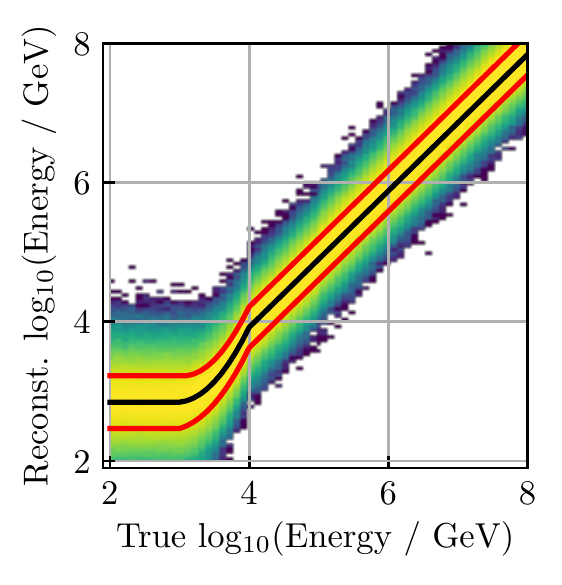}
   \caption{Energy reconstruction distribution as described by equation \ref{eq:reco_distr}. For low energies, which are relevant for this study, the distribution shows a bias towards higher energies. The x and y axis show the true and reconstructed particle energies respectively. The black line indicates the mean distribution value, while the red lines the 68-percentile.} 
   \label{fig:Reco_distr}
 \end{figure}

The IceCube detector is capable of reconstructing the direction from which a low-energy muon or stau comes with high accuracy. Given the symmetry of the Earth and of the detector, the information of interest about the direction is typically expressed in terms of the angle with respect to the zenith. For the energy range of interest, $E\in[$100 GeV, 1 TeV$]$, the resolution with which IceCube can reconstruct this angle is 1$^\circ$\cite{Aartsen:2014cva}, and this is folded into our simulations by smearing the particle arrival directions accordingly. The energy released in the detector's active volume is reconstructed with a resolution of ~20\% for muon events with energy above 1\,TeV. Below this threshold, the energy reconstruction is biased towards 700 GeV, see Figure \ref{fig:Reco_distr}.
The origin of this bias is related to the particles being minimally ionizing at these energies. For the energy estimation, current reconstruction methods rely on stochastic losses, which are negligible at low energies. Both the bias and resolution in the energy reconstruction are folded in our simulation. More details are given in \ref{sec:energy_reco}.

\begin{figure}[t]
   \centering
   \includegraphics[scale=1.]{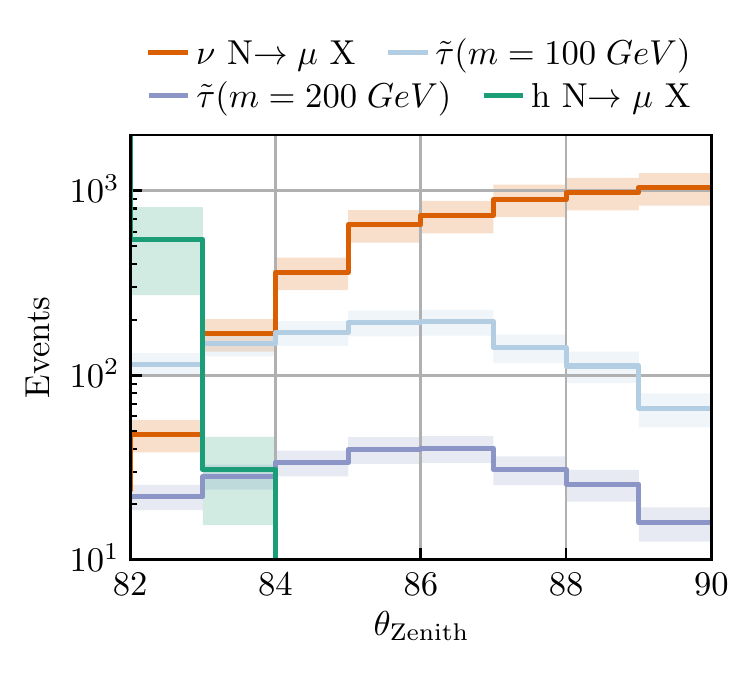}
   \caption{Predicted number of stau or muon events in IceCube after one year as a function of the arrival direction with respect to the zenith. For the stau events we have assumed Drell-Yan production and masses of 100 and 200 GeV, light blue and purple respectively.	For the muon events we show separately the contribution from hadronic interactions (green) and from neutrino interactions (orange). The shaded regions shows the model uncertainties.}
   \label{fig:Mass}
 \end{figure}

Figure~\ref{fig:Mass} shows the expected angular distribution for signal and background events in the public IceCube data set. Staus are barely stopped by the Earth or the material surrounding the detector and hence their angular distribution appears flat. However, because the amount of material to be traversed increases rapidly towards the horizon, their distribution is slightly peaked at 86$^\circ$. The muon contribution due to hadron interactions is maximal at the zenith (i.e. at 0$^\circ$) and decreases steeply for increasing angles as the muon-flux is attenuated by the ice-overburden over the detector. This contribution drops below the rate expected for staus towards the horizon, for angles above 84-85$^\circ$. On the contrary, the muon contribution due to neutrino interactions increases towards the horizon because that's where the flux of atmospheric neutrinos is largest. The distribution grows till about 90$^\circ$, after which it saturates. The opposite trend of the two background distributions creates a
small range of angles in which the background is minimal and the sensitivity to a stau signal is maximal. This range is independent of the stau mass, as shown in Figure \ref{fig:Mass}. Higher masses scale the distribution of the staus, while leaving the shape unchanged.

\begin{figure}[t]
   \centering
   \includegraphics[scale=1.]{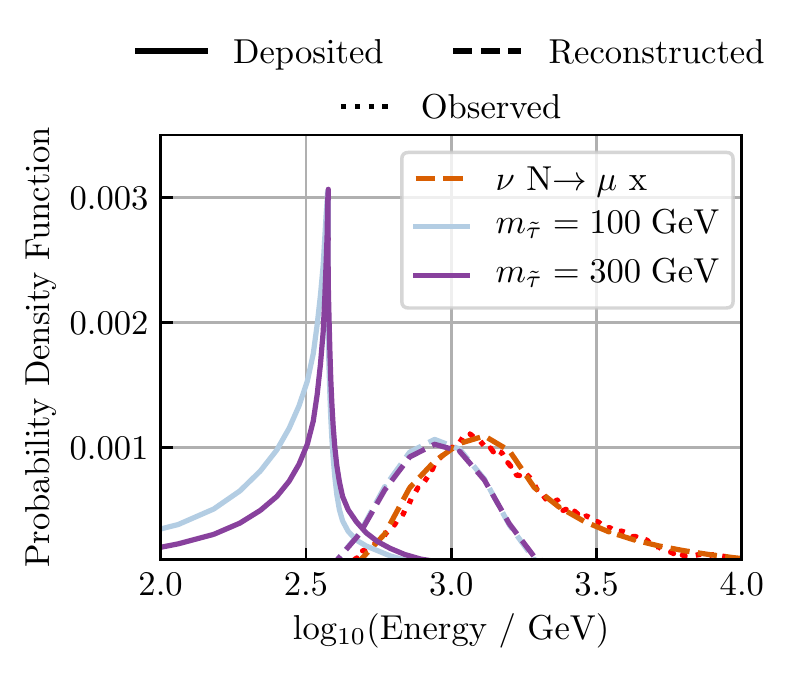}
   \caption{Energy distributions for the signal and background events. Shown in light blue and purple are the deposited energy distributions (solid line) and reconstructed energy distribution (dashed line) for the staus. In orange we show the background energy distribution as predicted (dashed line) and observed (dotted line). The Background distributions were scaled up by a factor of 5 to make the comparison with the signal prediction possible.}
   \label{fig:Energy_Distribution}
 \end{figure}

The energy distribution of the staus and muon background due to neutrinos for angles between 85 and 90$^\circ$ is shown in Figure~\ref{fig:Energy_Distribution}. 
As a consequence of the low rate of energy loss, staus deposit within the detector as much energy as as the lowest-energy muons. Their distribution is sharply peaked as they cross the entire detector depositing always the same energy regardless of their kinetic energy. The full distribution is in the energy range for which the energy reconstruction is biased and the events are given a random energy estimation that is independent by their original energy distribution. After the energy reconstruction, there is a minimal difference between staus of different masses. The reconstructed energy distribution for the muon background partially overlaps with the stau distribution but has a tail extending towards high energies. The different energy distribution provides another handle to separate our signal
from the background.

\begin{figure}[t]
   \centering
   \includegraphics[scale=1.]{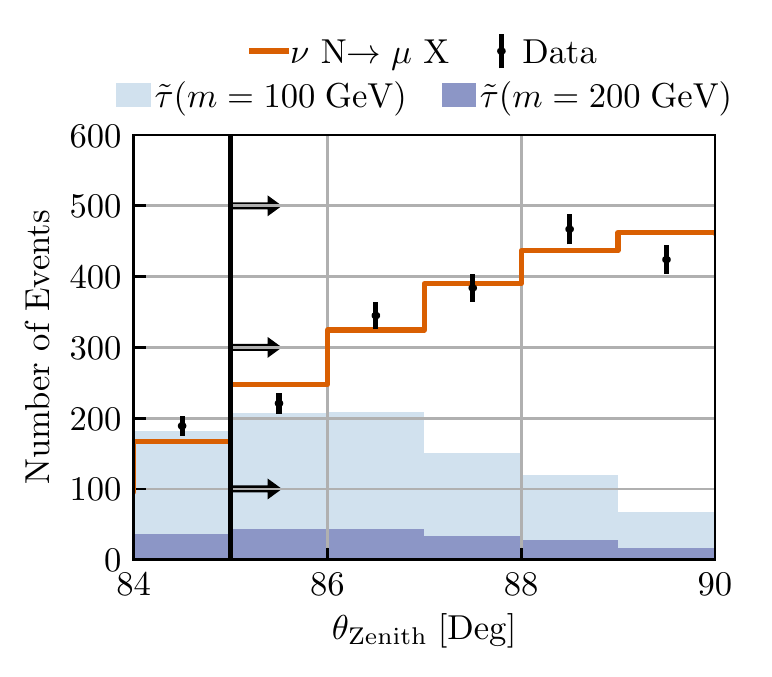}
   \caption{The expected number of events in a background-only model (orange) and staus of different mass (blue) after imposing the energy cut $E\in [100\;\mathrm{GeV}, 1\;\mathrm{TeV}]$, for $\theta_{\rm Zenith}>85^\circ$ and including reconstruction effects. The measured number of track events in IceCube is shown, including statistical errors. The plot was generated using QGSJET and H3a as the interaction and primary models respectively. The solid black lines and arrows emphasizes the region used in this analysis.}
   \label{fig:Counts}
 \end{figure}

Based on our modeling, we identified the observable space in which the ratio between the stau signal and the muon-induced background is maximal. This space
corresponds to angles between 85 and 90$^\circ$ and energies between 0.1 and 1\,TeV. The most efficient way to extract the signal would be to perform a bivariate analysis
in energy and angle. Such an analysis should take into account the systematic uncertainties related to the detector response, in particular related to the bias in the energy reconstruction. Given that these uncertainties are not completely available in literature, we opted not to use in the analysis the full shape of the energy distribution but
only to apply a loose cut to remove the high-energy part of the background events ($E>1$\,TeV). This approach reduces the sensitivity of the analysis but makes it more
robust. We applied these simple selection criteria ($100\;\mathrm{GeV} < E < 1\;\mathrm{TeV}$ and $85^\circ < \theta_\mathrm{Zenith} < 90^\circ$)to the data of IceCube ~\cite{IceCubeData2018, Aartsen:2016oji}. It follows well the background-only distribution. The angular distribution for events with energy between 0.1 and 1\,TeV is shown in Figure~\ref{fig:Counts}.
Also, in this case the data points follow well the background-only distribution. The p-value of our data given the background-only hypothesis is 0.1, supporting
our intuition that there is no evidence for a signal. Assuming the currently leading experimental limit on the stau mass ($m_{\tilde\tau}=$430\,GeV), we expect to retain 8 stau events. To extract a limit on the mass of the staus we perform a binned likelihood fit of the data shown in Figure~\ref{fig:Counts}. The fit has one free parameter, the mass of the stau particle. The rates of the background components are fixed by our modeling. The best fit results in 0 stau events. Inverting a standard frequentist hypothesis test \cite{Feldman:1997qc},  we set a lower bound on the stau mass of $m_{\tilde{\tau}}> 320\;\mathrm{GeV}$ at 95\% C.L. This lies approximately 10\% above the expected limit from simulations.

This result does not include systematic uncertainties. Because of the way we constructed the analysis -- i.e. by using only a loose cut on the energy and a bin size equal to the angular resolution of the experiment -- we expect systematic uncertainties to be negligible, to first approximation. Future analysis based on larger data samples that exploit fully the energy information will need to refine the treatment of the systematic uncertainties.

\begin{figure}[htb]
   \centering
   \includegraphics[scale=1.]{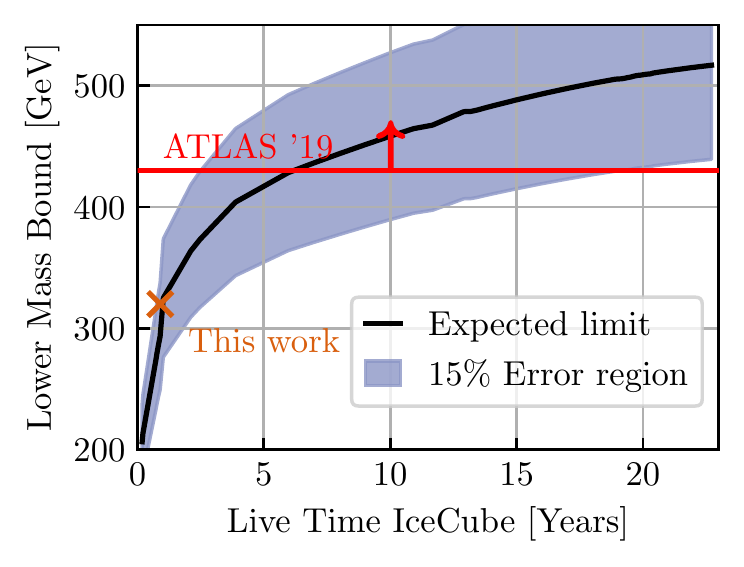}
   \caption{Projection of the sensitivity to long-lived staus at IceCube as a function of time, assuming Drell-Yan production. The red line corresponds to the current ATLAS limit \cite{Aaboud:2019trc}. The orange cross shows the here obtained limit, while the black line plots the expected one. The purple shaded region shows the 15\% uncertainty region.} 
   \label{fig:Limit_Capabilities}
 \end{figure}

Our limit of $m_{\tilde{\tau}}> 320\;\mathrm{GeV}$ is the most stringent constraint on stau masses ever set by a non collider experiment. This proves the potential of our analysis approach. The analysis can be improved significantly by increasing the exposure (see Figure~\ref{fig:Limit_Capabilities}). With the ten years of IceCube data that have been already recorded, and assuming no improvements to the reconstruction, simulation or energy resolution, our analysis approach would provide a sensitivity to stau masses of up to 450\,GeV. Thus, neutrino telescopes have the chance to probe an unexplored parameter space, together with future LHC experiments but using a very different and complementary technique.

Furthermore, several neutrino telescopes, such as P-ONE \cite{Agostini:2020aar}, KM3NET \cite{Adrian-Martinez:2016fdl}, GVD \cite{Avrorin:2017tse} and IceeCube-Gen2 \cite{vanSanten:2017chb}, are currently taking data or are in preparation. While their designs are different, they are all based on the detection of the Cherenkov light emitted by charged particles traveling through water or ice. The analysis performed in this work using data from IceCube can be extended to these upcoming datasets as well. Finally, this search strategy is not limited to the stau particle but can be utilized to search for other long-lived, charged particles beyond the Standard Model.

 \section{Acknowledgements}
 We thank Martin Wolf and Spencer Klein for the fruitful discussions.
 The material presented in this publication is based
 upon work supported by the Sonderforschungsbereich Neutrinos and Dark Matter in
 Astro- and Particle Physics (SFB1258). It was also funded by the Deutsche
 Forschungsgemeinschaft (DFG, German Research Foundation) under Germany's
 Excellence Strategy – EXC-2094 – 390783311. M.A. is supported by the Science and Technology Facilities Council 
(STFC) Grant No. ST/T004169/1.

 \appendix
 \section{Atmospheric shower Simulation}\label{sec:app:atmos_shower}
 The particle interactions are modeled with Sibyll 2.3c \cite{Riehn:2017mfm}, EPOS-LHC \cite{Pierog:2013ria}, QGSJET-II \cite{Ostapchenko:2010vb} and DPMJET-III \cite{Roesler:2000he}. For the cosmic ray models we use the Gaisser-Hillas models H3a and H4a \cite{Gaisser:2011cc} and Gaisser-Stanev-Tilav Gen 3 and 4 \cite{Gaisser:2013bla}. The resulting differences are used as an estimate of the uncertainties in our calculation. We found these to be negligible compared to other uncertainties in our analysis that will be discussed later on. We use the NRLMSISE-00 \cite{Hedin:1991grsp, Picone:2002grsp} model to simulate the atmosphere.

 \section{Stau production}\label{sec:app:stau_cross}
 We use the built in MSSM model in MadGraph. We furthermore adopt the parton distribution functions (PDF) from LHAPDF6 \cite{Buckley:2014ana}. Concretely, we use the \texttt{CT10nlo} \cite{Lai:2010vv} PDF as well as the \texttt{NNPDF30\_nnlo\_nf\_5\_pdfas} from the NNPDF3.0 \cite{Ball:2014uwa} PDF set. This allows us to include the uncertainties due to different PDF parametrizations. For the interactions of other hadrons with air, we scale the $p-p$ cross-sections following \cite{Ahlers:2007js}.

 \section{Effective areas}\label{sec:app:effective_areas}
 To calculate the contribution of neutrino-induced muons, we fold the 2D effective area, as a function of energy and declination, from \cite{Aartsen:2016oji} with the neutrino fluxes. The neutrino energy to muon energy mapping is approximated using the normalized 3D effective areas given in \cite{Aartsen:2015rwa}.
 \\
 To make predictions for the stau component, we require a detector response to staus. We use the same approach as for the muons and divide the convolution by the total neutrino cross-section. The resulting efficiency includes effects of muon propagation in the ice. These we compensate by scaling the results, so at 1 TeV the effective area for muons corresponds to the spatial area of the detector, $10^6\;\mathrm{m}^2$. This results in a signal efficiency of 78\%.

 \section{Energy Reconstruction}\label{sec:energy_reco}
 We include the effects of energy reconstruction as described in \cite{Aartsen:2013vja, Aartsen:2015rwa} by constructing a function, mapping the true particle energy, $E_\mathrm{true}$, to the reconstructed energy, $E_\mathrm{reco}$, of the form
 \begin{equation}\label{eq:reco_distr}
   \begin{split}
     E_\mathrm{reco} & =  \\
     & \left\{\begin{array}{ll}
       \text{LogNorm}(E_\mathrm{true}, \sigma_1, \mu_1) &  E_\mathrm{true} < 1\;\mathrm{TeV}\\
       \text{Linear interpolation} & E_\mathrm{true} \in [1, 10]\;\mathrm{TeV}\\
       \text{LogNorm}(E_\mathrm{true}, \sigma_2, \mu_2) &  E_\mathrm{true} > 10\;\mathrm{TeV}
     \end{array}\right. .
   \end{split}
 \end{equation}
 To get the specific values for $\sigma$ and $\mu$, we fitted the background prediction to the data for energies above 1 TeV. For such energies we do not expect any stau events. The results are $\sigma_1 = 0.4$, $\mu_1 = 700\;\mathrm{GeV}$, $\sigma_2 = 0.3$ and $\mu = E_\mathrm{true}$. The energy reconstruction distribution is shown in Figure (\ref{fig:Energy_Distribution}). These values agree well with those shown in \cite{Aartsen:2013vja}. To map the stau energies to their reconstructed energies, we map their energy deposit to muon energies with an equivalent loss according to equation \ref{eq:energyloss} and then proceed as with the muons.
 \nocite{*}
\bibliographystyle{elsarticle-num}
 \bibliography{Staus}

\end{document}